\documentclass[letterpaper, 10 pt, conference]{ieeeconf}
\usepackage{amsfonts,amsmath,amssymb} % assumes amsmath package installed

\usepackage{psfrag,color}
\usepackage{enumerate,cite,latexsym,graphicx,subfig}
\newtheorem{theorem}{Theorem}
\newtheorem{lemma}{Lemma}

\newtheorem{definition}{Definition}

\title{\LARGE \bf Physical Interpretations of Negative Imaginary Systems Theory}

\author{Ian R.~Petersen %
\thanks{This work was supported by the
Australian Research Council (ARC).}%
\thanks{Ian R. Petersen is with the School of  Engineering and Information Technology, 
        University of New South Wales at the Australian Defence Force Academy, Canberra ACT 2600, Australia.
         {\tt\small i.r.petersen@gmail.com} } 
}%
        
\def\begce{\begin{center}}
\def\endce{\end{center}}
\def\begar{\begin{array}}
\def\endar{\end{array}}
\def\begeq{\begin{equation}}
\def\endeq{\end{equation}}
\def\begdi{\begin{displaymath}}
\def\enddi{\end{displaymath}}
\def\begdis{\begin{eqnarray*}}
\def\enddis{\end{eqnarray*}}
\def\begeqa{\begin{eqnarray}}
\def\endeqa{\end{eqnarray}}

\begin{document}

\maketitle
\thispagestyle{empty}
\pagestyle{empty}

\begin{abstract}
This paper presents some physical interpretations of recent stability results on the feedback interconnection of negative imaginary systems. These interpretations involve spring mass damper systems coupled together by springs or RLC electrical networks coupled together via inductors or capacitors. 
\end{abstract}

\section{Introduction} \label{sec:intro}
In recent years the theory of negative imaginary systems has emerged as a useful complement to positive real theory and passivity theory; e.g., see \cite{LP06,LP_CSM,ANG06}. It is well known that linear mechanical systems with force inputs and collocated velocity outputs lead to positive real transfer functions; e.g., see \cite{{BLME07}}. Similarly, linear mechanical systems with force inputs and collocated position outputs lead to negative imaginary transfer functions; e.g., see \cite{LP_CSM}. In this paper we extend these ideas to consider mechanical interpretations of the main negative imaginary stability results and compare them with corresponding interpretations of positive real stability results. The negative imaginary stability results involve the positive feedback interconnection between two negative imaginary system, one of which is regarded as the controller and one of which is regarded as the plant; e.g., see \cite{XiPL1,MKPL10}.  In our mechanical interpretation of these results, the plant is a spring mass damper system and the controller is another spring mass damper system which is coupled to it via a spring. This is similar to the behavioural systems theory notion of control by interconnection; e.g., see \cite{PW98}. This paper also presents similar RLC electrical circuit interpretations of the negative imaginary stability results in which the coupling between the plant RLC circuit and the controller RLC circuit is via capacitive or inductive coupling. 

The mechanical interpretations presented in this paper may help to better understand the negative imaginary stability results and motivate extensions to these results. Also, the mechanical interpretation of the controller may be useful in controller design and tuning. 

\section{Negative Imaginary and Positive Real Systems Theory}
\label{sec:NI_systems}

\noindent
{\bf Negative imaginary and positive real systems.}
Stability results about the positive feedback interconnection of negative imaginary systems take their simplest form in the case in which none of the systems have poles at the origin. Hence, we will consider this case first along with a stardard positive real stability result. In the sequel, we will look at the more general case in which poles at the origin are allowed.

\begin{definition}[See \cite{XiPL1}]
  \label{D1}
  A square real-rational proper transfer function matrix $G(s)$ is
  termed \emph{negative imaginary} (NI) if
  \begin{enumerate}
  \item $G(s)$ has no poles at the origin and in $\Re[s]>0$;
  \item $j[R(j\omega)-R^*(j\omega)]\ge0$ for all $\omega\in(0,\infty)$ except
    values of $\omega$ where $j\omega$ is a pole of $G(s)$;
  \item If $j\omega_0$, $\omega_{0}\in(0,\infty)$, is a pole of $G(s)$, it is
    at most a simple pole, and the residue matrix $K_0\triangleq\lim_{s\to
      j\omega_0}(s-j\omega_0)jG(s)$ is positive semidefinite Hermitian.
  \end{enumerate}
\end{definition}

\begin{definition}[See \cite{LP06}]
  \label{D2}
  A square real-rational proper transfer function matrix $G(s)$ is
  termed \emph{strictly negative imaginary} (SNI) if
  \begin{enumerate}
  \item $G(s)$ has no poles in $\Re[s]\ge0$;
  \item $j[G(j\omega)-G^*(j\omega)]>0$ for $\omega\in(0,\infty)$.
  \end{enumerate}
\end{definition}

In addition to looking at physical interpretations of negative imaginary stability results, we will also compare these interpretations with physical interpretations of positive real stability results. Hence, we introduce  the corresponding notions of positive real and weakly strictly positive real transfer functions.

\begin{definition}[See \cite{AV73}]
  \label{D3}
  A square real-rational transfer function matrix $G(s)$ is termed
  \emph{positive real} (PR) if
  \begin{enumerate}
  \item No element of $G(s)$ has a pole in $\Re[s]>0$;
  \item $G(s)+ G^{*}(s) \ge 0$ for $\Re[s]>0$.
  \end{enumerate}
\end{definition}

\begin{definition}[See \cite{BLME07}]
  \label{D4}
  A non-zero square real-rational transfer function matrix $G(s)$ is
  \emph{weakly strictly positive real} (WSPR)
if 
\begin{enumerate}
  \item $G(s)$ has no poles in $\Re[s]\ge0$;
  \item $G(j\omega)+G^*(j\omega)>0$ for $\omega\in(-\infty,\infty)$.
\end{enumerate}
\end{definition}

The following lemma makes clearer the relationship between  negative imaginary and positive real transfer functions. 

\begin{lemma}[Theorem 2.7.2 of \cite{AV73}]
  \label{L1}
  Let $G(s)$ be a square real-rational transfer function matrix. Then
  $G(s)$ is positive real if and only if
  \begin{enumerate}
  \item No element of $G(s)$ has a pole in $\Re[s]>0$;
  \item $G(j\omega)+G^{*}(j\omega)\ge0$ for all real $\omega$ except values of
    $\omega$ where $j\omega$ is a pole of $G(s)$;
  \item If $j\omega_{0}$ is a pole of any element of $G(s)$, it is at most a
    simple pole, and the residue matrix, $K_{0}\triangleq\lim_{s\to
      j\omega_{0}}(s-j\omega_{0})G(s)$ in case $\omega_{0}$ is finite, and
    $K_{\infty}\triangleq\lim_{\omega\to\infty}\frac{G(j\omega)}{j\omega}$ in case
    $\omega_{0}$ is infinite, is positive semidefinite Hermitian.
  \end{enumerate}
\end{lemma}

\begin{figure}[tbh]
  \centering
  \includegraphics[width=7cm]{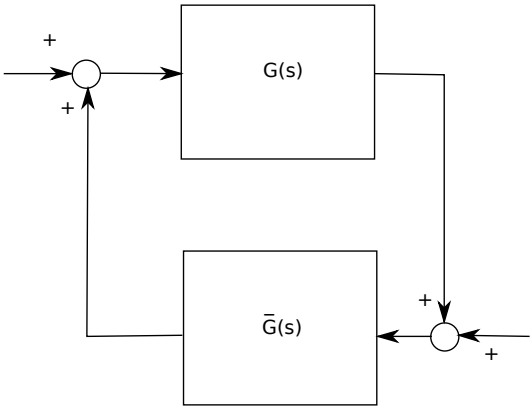}
  \caption{Positive feedback interconnection}
  \label{F1}
\end{figure}

Now we recall a  stability result for the positive feedback interconnection 
 of two negative imaginary systems, denoted by
$[G(s),\bar G(s)]$, as shown in Figure \ref{F1}.

\begin{theorem}[See \cite{XiPL1}]
  \label{T1}
  Given a negative imaginary transfer function matrix $G(s)$ and a strictly
  negative imaginary transfer function matrix $\bar G(s)$ that also satisfy
  $G(\infty)\bar G(\infty)=0$ and $\bar G(\infty)\ge0$. Then the positive feedback
  interconnection $[G(s),\bar G(s)]$ is internally stable if and only if
  \begin{equation}
\label{DC_gain}
\lambda_{\mathrm{max}}(G(0)\bar G(0)) < 1.
\end{equation}
\end{theorem}

\begin{figure}[tbh]
  \centering
  \includegraphics[width=7cm]{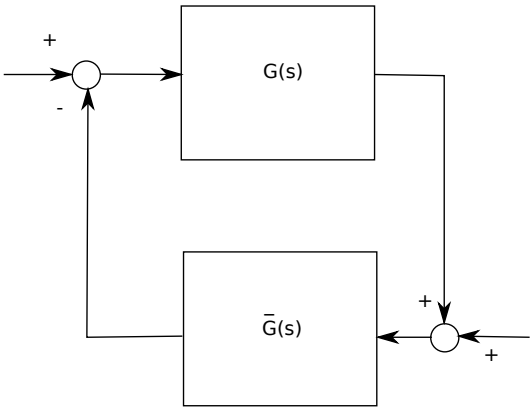}
  \caption{Negative feedback interconnection}
  \label{F2}
\end{figure}

Also, we recall a  stability result for the negative feedback interconnection 
 of two positive real systems, denoted by
$[G(s),-\bar G(s)]$, as shown in Figure \ref{F2}.

\begin{theorem}[See for example Lemma 3.37 in \cite{BLME07}]
  \label{T2}
  Given a positive real transfer function matrix $G(s)$ and a weakly strictly
  positive real transfer function matrix $\bar G(s)$ such that their negative feedback interconnection is well posed. Then the negative feedback
  interconnection $[G(s),-\bar G(s)]$ is internally stable.
\end{theorem}

\noindent
{\bf Negative imaginary systems with poles at the origin.}
We now recall some recent stability results for the feedback interconnection between negative imaginary systems containing poles at the origin. This situation may arise in control systems in which the plant has free body motion or the controller contains integral action. The complexity of these results depends on the special cases being considered. For the purposes of this paper, we will present two of the simplest results. 

We begin with a generalized definition of the negative imaginary property.

\begin{definition}[See \cite{MKPL10}]\label{Def:NI}
A square transfer function matrix $G(s)$ is NI if the following
conditions are satisfied:
\begin{enumerate}
\item $G(s)$ has no pole in $\Re[s]>0$.
\item For all $\omega >0$ such that $j\omega$ is not a pole of $G(s
)$,
$
    j\left( G(j\omega )-G(j\omega )^{\ast }\right) \geq 0.
$
\item If $s=j\omega _{0}$ with $\omega _{0}>0$ is a pole of $G(s)$, then it is a simple pole and the residue matrix $K=\underset{%
s\longrightarrow j\omega _{0}}{\lim }(s-j\omega _{0})jG(s)$ is
  positive semidefinite.
 \item If $s=0$ is a pole of $G(s)$, then
$\underset{s\longrightarrow 0}{\lim }s^{k}G(s)=0$ for all $k\geq3$
and $\underset{s\longrightarrow 0}{\lim }s^{2}G(s)$ is 
positive semidefinite.
\end{enumerate}
\end{definition}

Then, we define the following matrices which will be used in the stability conditions to be presented:
\begin{align}
G_2&=\underset{s\longrightarrow 0}{\lim }s^2G(s),\;\;
G_1=\underset{s\longrightarrow 0}{\lim }s \left(G(s)-\frac{G_2}{s^2}\right).
\label{f_G0}
\end{align}

Roughly speaking, transfer function matrices $G(s)$ with only single poles at the origin have $G_2 =0$ and transfer function matrices with only double poles at the origin have $G_1 =0$. 

\begin{theorem}[See \cite{MKPL10}]
\label{T3}
Suppose that the transfer function matrix  $\bar{G}(s)$ is SNI and
the strictly proper transfer function matrix  $G(s)$ is  NI with
$G_2=0$ and $G_1$ invertible. Then,
the closed-loop positive-feedback interconnection  $[G(s),\bar{G}(s)]$ is internally stable if and only if 
\begin{equation}
\label{DC_gain1}
\bar{G}(0)<0.
\end{equation}
\end{theorem}

\begin{theorem}[See \cite{MKPL10}]
\label{T4}
Suppose that the transfer function matrix  $\bar{G}(s)$ is SNI and
the strictly proper transfer function matrix  $G(s)$ is  NI with
$G_1=0$ and $G_2>0$. Then,
the closed-loop positive-feedback interconnection between $[G(s),\bar{G}(s)]$ is internally stable if and only if 
\begin{equation}
\label{DC_gain2}
\bar{G}(0)<0.
\end{equation}
\end{theorem}

\section{Mass Spring Damper System Interpretations}
\label{sec:MSD_systems}
\begin{figure}[tbh]
  \centering
  \includegraphics[width=7cm]{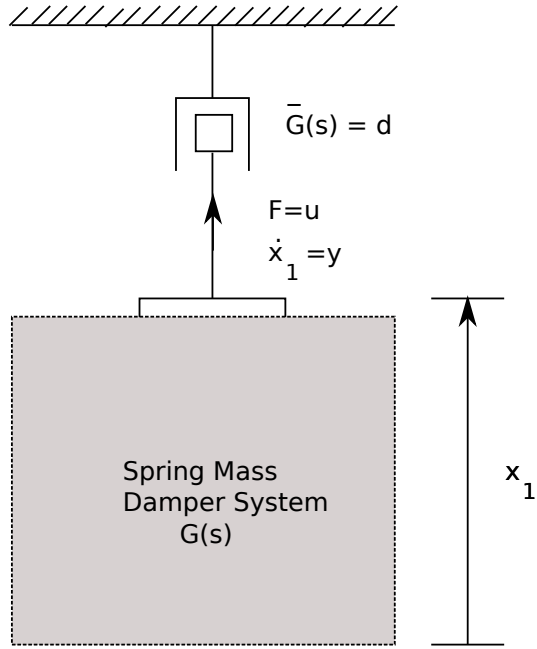}
  \caption{Spring mass damper interpretation of the PR stability theorem}
  \label{F3}
\end{figure}

\begin{figure}[tbh]
  \centering
  \includegraphics[width=7cm]{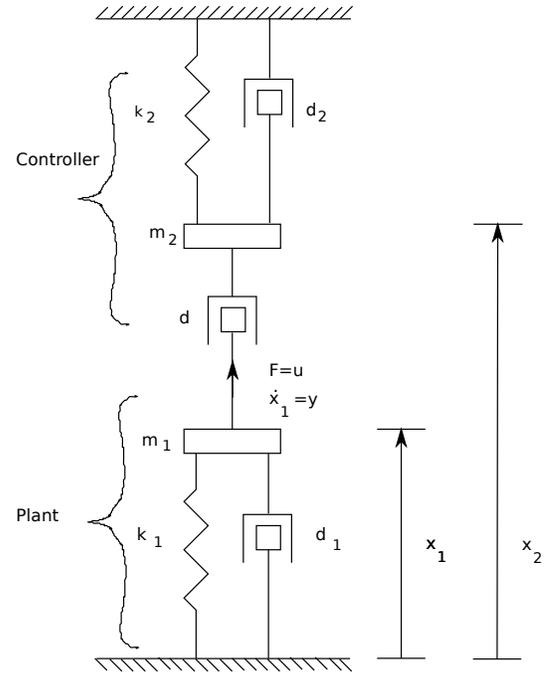}
  \caption{Spring Mass Damper illustration of the PR stability theorem with dynamic plant and controller}
  \label{F4}
\end{figure}

We first consider a physical interpretation of  Theorem \ref{T2} for the case in which the controller transfer function $\bar G(s) = d > 0$ is a constant  and the plant transfer function $G(s)$ is a scalar transfer function corresponding to a spring mass damper system in which the plant input is the force applied to a given mass and the plant output is the corresponding velocity of that mass. The equations of motion for such a plant can be written in the form
\begin{eqnarray*}
M\ddot x + D \dot x + K x &=& Lu,~~
y = L^T \dot x
\end{eqnarray*}
where $M > 0$, $D \geq 0$, $K > 0$, $x \in\mathbb{R}^n$ corresponds to the vector of mass positions,  $u \in \mathbb{R}$ corresponds to the force input and $y \in \mathbb{R}$ corresponds to the velocity output; e.g., see \cite{BLME07,PRE11}. The corresponding plant transfer function is then
$
G(s) = sL^T\left(s^2M+sD+K\right)^{-1}L.
$
It is straightforward to verify that $G(s)$ is PR and $\bar G(s)$ is WSPR. In this case, the negative feedback interconnection between $G(s)$ and $\bar G(s)$ can be interpreted as shown in Figure \ref{F3}. 
Thus in this case, the positive real stability theorem, Theorem \ref{T2} can be interpreted as saying that the addition of a damper to a spring mass system will lead to the overall system being internally stable. We can think of the damper as a static velocity feedback controller; e.g., see \cite{BAL79}. This interpretation can be extended to allow for a more general dynamic controller with a WSPR controller transfer function $\bar G(s)$ as illustrated in Figure \ref{F4}.

From this figure, we can write down the equations of motion as follows:
\begin{eqnarray*}
m_1 \ddot x_1 &=& F - k_1 x_1 - d_1 \dot x_1, \nonumber \\
m_2 \ddot x_2 &=& -k_2 x_2 - d_2\dot x_2 -  d(\dot x_2 - \dot x_1).
\end{eqnarray*}
The input to the plant is the force $F$ and the output of the plant is the velocity $\dot x_1$. Taking Laplace transforms, we obtain the plant transfer function
$
G(s) = \frac{s}{m_1s^2+d_1s+k_1}.
$
If we assume $m_1 > 0$, $d_1 \geq 0$ and $k_1 > 0$, it is straightforward to verify that $G(s)$ is PR. 
For a negative feedback interconnection, the input of the controller is the output of the plant $\dot x_1$ and the output of the controller is minus the input of the plant $-F = -d (\dot x_2 - \dot x_1)$. Taking Laplace transforms, we obtain the controller transfer function
$
\bar G(s) = d \frac{m_2 s^2 + d_2s + k_2}{m_2 s^2 + (d+d_2)s + k_2}
$
which is WSPR for $m_2 > 0$, $d>0$, $d_2 >  0$ and $k_2 > 0$. Thus, the PR stability result in this case can be interpreted as simply ensuring the stability of the complete spring mass system shown in Figure \ref{F4}. The fact that the PR stability result can be used to ensure stability is directly related to the fact that the plant subsystem is coupled to the controller subsystem via a damper for which the force is proportional to velocity. Note that in this example, the condition $d_2 >  0$ is required for the controller transfer function $\bar G(s)$ to be WSPR. If $d_2 =  0$ then $\bar G(s)$ will have a pair of complex zeros on the imaginary axis. Also, $G(s)$ is never WSPR since it always has a zero at the origin. 

\begin{figure}[tbh]
  \centering
  \includegraphics[width=7cm]{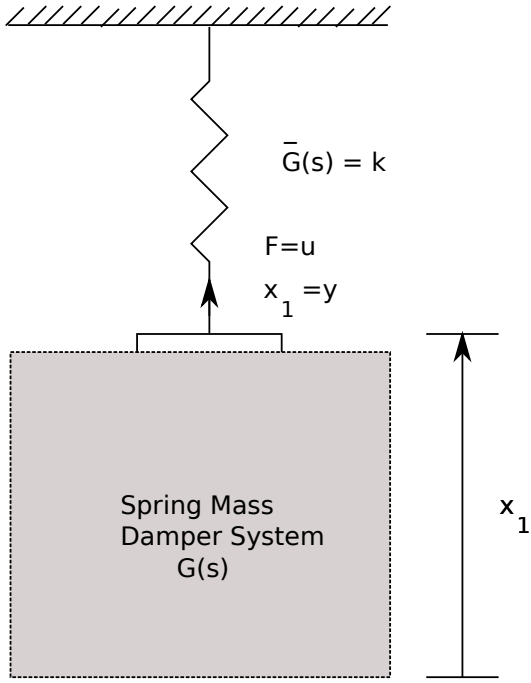}
  \caption{Spring mass damper illustration of the NI stability theorem}
  \label{F5}
\end{figure}

We will now see that the NI stability result, Theorem \ref{T1}, corresponds to the case in which the plant subsystem is coupled to the controller subsystem via a spring in which the force is proportional to displacement. Indeed,  consider a physical interpretation of the Theorem \ref{T1} for the case in which the controller transfer function  $\bar G(s) = -k < 0$  is a constant and the plant transfer function $G(s)$ is a scalar transfer function corresponding to a spring mass damper system in which the plant input is the force applied to a given mass and the output is the corresponding displacement of that mass. The equations of motion for such a system can be written in the form
\begin{eqnarray*}
M\ddot x + D \dot x + K x &=& Lu,~~
y = L^T  x
\end{eqnarray*}
where $M > 0$, $D > 0$, $K > 0$, $x \in\mathbb{R}^n$ corresponds to the vector of mass positions,  $u \in \mathbb{R}$ corresponds to the force input and $y \in \mathbb{R}$ corresponds to the displacement output; e.g., see \cite{BLME07,PRE11}. The corresponding transfer function is then
$
G(s) = L^T\left(s^2M+sD+K\right)^{-1}L.
$
It is straightforward to verify that $G(s)$ is SNI and $\bar G(s)$ is NI. In this case, the positive feedback interconnection between $G(s)$ and $\bar G(s)$ can be interpreted as shown in Figure \ref{F5}. 
Note that in this case, we have reversed the strict and non strict transfer functions compared to the PR case. Also, in this case, we do not actually need to assume that $k > 0$ in order to ensure stability. Indeed according to Theorem \ref{T1}, we only require the DC gain condition (\ref{DC_gain}) which is
$
G(0) \bar G(0) = -L^T K L k < 1.
$
This will be automatically satisfied if $k > 0$. However, it will also be satisfied if $k < 0$ and 
$
-k < \frac{1}{L^T K L}. 
$
That is, the NI stability theorem can allow for the situation in which the controller corresponds to a negative spring provided it is not too large. 

\begin{figure}[tbh]
  \centering
  \includegraphics[width=7cm]{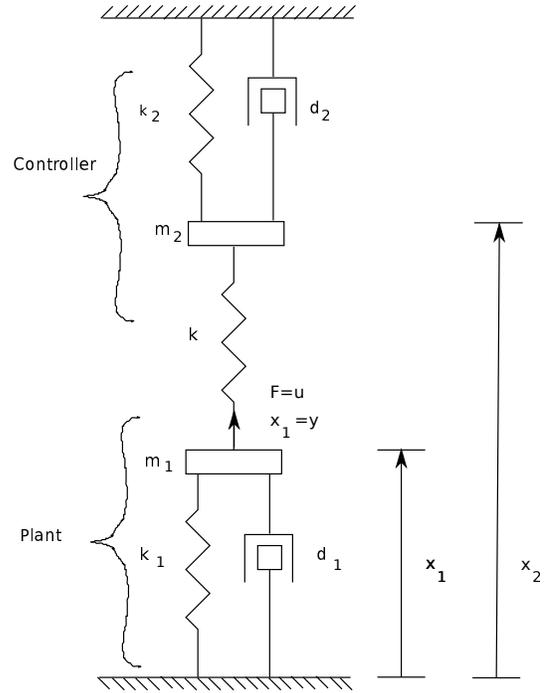}
  \caption{Spring mass damper illustration of the NI stability theorem with dynamic plant and controller}
  \label{F6}
\end{figure}

From the above example, we can see that the application of the NI stability result Theorem \ref{T1} arises when the plant is coupled to the controller via a spring. 
This interpretation can be extended to allow for a more general dynamic controller with an NI transfer function $\bar G(s)$ as illustrated in Figure \ref{F6}.

From this figure, we can write down the equations of motion as follows:
\begin{eqnarray*}
m_1 \ddot x_1 &=& F - k_1 x_1 - d_1 \dot x_1 \nonumber \\
m_2 \ddot x_2 &=& -k_2 x_2 - d_2\dot x_2 -  k( x_2 -  x_1).
\end{eqnarray*}
The input of the plant is the force $F$ and the output of the plant is the displacement $ x_1$. Taking Laplace transforms, we obtain the plant transfer function
$
\bar G(s) = \frac{1}{m_1s^2+d_1s+k_1}.
$
If we assume $m_1 > 0$, $d_1 >  0$ and $k_1 > 0$ it is straightforward to verify that $\bar G(s)$ is SNI. 
For a positive feedback interconnection, the input of the controller is the output of the plant $x_1$ and the output of the controller is  the input of the plant $F = k ( x_2 -  x_1)$. Taking Laplace transforms, we obtain the controller transfer function
$
 G(s) = -k \frac{m_2 s^2 + d_2s + k_2}{m_2 s^2 + d_2s + k+k_2}
$
which is NI for $m_2 > 0$,  $d_2 \geq  0$  and $k+k_2 \geq 0$. Thus, the NI stability result in this case can be interpreted as simply ensuring the stability of the complete spring mass system shown in Figure \ref{F6}. The fact that the NI stability result can be used to ensure stability is directly related to the fact that the plant subsystem is coupled to the controller subsystem via a spring for which the force is proportional to displacement. 

Note that in this case, since the controller is only required to be NI, we can allow $d_2 =  0$ and also, we can allow $k$ or $k_2$ to be negative provided that $k+k_2 \geq 0$ and the DC gain condition (\ref{DC_gain}) is satisfied. For this example, the DC gain condition (\ref{DC_gain}) is 
$
\bar G(0) G(0) = \frac{1}{k_1}(-\frac{k k_2}{k+k_2}) < 1.
$
If both $k$ and $k_2$ are positive, this condition will be automatically satisfied. However, if one of $k$ and $k_2$ is negative, this condition will require an additional restriction of
$
\frac{1}{k}+\frac{1}{k_2} < -\frac{1}{k_1}.
$
In this case, this condition will in fact imply that the condition $k+k_2 \geq 0$ is satisfied. 

A distinction between the use of the PR stability theorem and the NI stability theorem is that the PR stability theorem requires that all components of the controller to be passive whereas the NI stability theorem allows for non-passive spring components in the controller. However, since the application of the NI stability theorem requires that the plant be SNI this means that in this example, the plant damping $d_1$ was required to be positive whereas in the the application of the PR stability theorem, the plant damping could be zero. 

\begin{figure}[tbh]
  \centering
  \includegraphics[width=7cm]{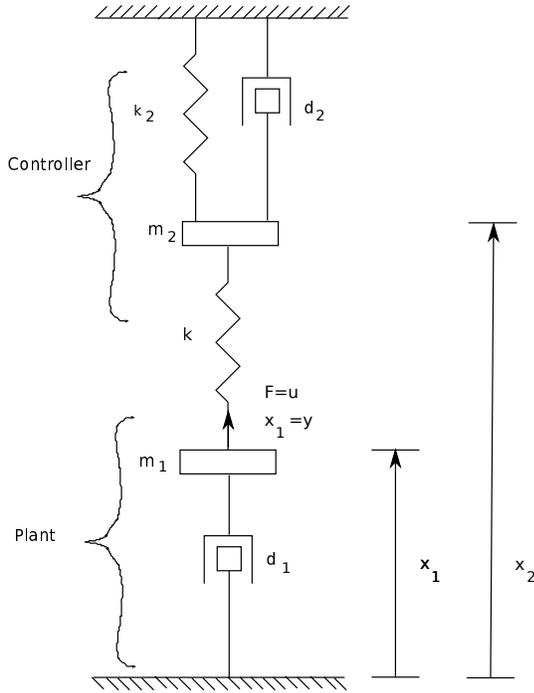}
  \caption{Spring mass damper illustration of the NI stability theorem with free body plant and a dynamic controller}
  \label{F7}
\end{figure}

We now consider a mass spring damper interpretation of Theorems \ref{T3} and \ref{T4} corresponding to a plant transfer function with poles at the origin. In our mass spring damper interpretation, this will correspond to a plant with free body motion such as shown in Figure \ref{F7}. 

From this figure, we can write down the equations of motion as follows:
\begin{eqnarray*}
m_1 \ddot x_1 &=& F  - d_1 \dot x_1, \nonumber \\
m_2 \ddot x_2 &=& -k_2 x_2 - d_2\dot x_2 -  k( x_2 -  x_1).
\end{eqnarray*}
The input of the plant is the force $F$ and the output of the plant is the displacement $ x_1$. Taking Laplace transforms, we obtain the plant transfer function
\begin{equation}
\label{G7}
G(s) = \frac{1}{m_1s^2+d_1s}.
\end{equation}
If we assume $m_1 > 0$ and $d_1 \geq  0$ it is straightforward to verify that $G(s)$ is NI according to Definition \ref{Def:NI}. 
For a positive feedback interconnection, the input of the controller is the output of the plant $x_1$ and the output of the controller is  the input of the plant $F = k ( x_2 -  x_1)$. Taking Laplace transforms, we obtain the controller transfer function
$
\bar G(s) = -k \frac{m_2 s^2 + d_2s + k_2}{m_2 s^2 + d_2s + k+k_2}
$
which is SNI for $m_2 > 0$,  $d_2 >  0$, $k \neq 0$, $k_2 \neq 0$  and $k+k_2 > 0$. 

If we assume $d_1 > 0$, then the conditions of Theorem \ref{T3} will be satisfied provided the  DC gain condition (\ref{DC_gain1}) holds: 
\begin{equation}
\label{DC7}
\bar G(0) = -\frac{k k_2}{k+k_2} < 0.
\end{equation}
This condition will be satisfied if $k > 0$ and $k_2 >0$ and in fact it can only be satisfied if $k > 0$ and $k_2 >0$. Thus, in this case, we cannot allow the use of active components in the controller. 

Note that this example can also illustrate the use of Theorem \ref{T3} when the controller includes integral action. In this case, we reverse the role of the plant and the controller in Figure \ref{F7}. In this case, the plant is a spring mass damper system which is coupled via a spring such that the input to the plant is the displacement of the string and the output of the plant is the corresponding force provided by the spring. Also, the controller with integral action is given as in (\ref{G7}). 

In the case that $d_1= 0$ then the conditions of Theorem \ref{T4} will be satisfied provided the same DC gain condition (\ref{DC7}) is satisfied. This corresponds to the case in which the plant transfer function has a double pole at the origin. 

Note that in this section, we have restricted attention to spring mass damper systems operating in one dimension and such that all transfer functions are SISO. However, it would be straightforward to construct similar examples involving spring mass damper systems operating in two or three dimensions. Also, it would be straightforward to construct examples involving multiple spring couplings between the plant and controller which would correspond to MIMO transfer functions.

\section{Electrical Circuit Interpretations}
\label{sec:RLC_systems}
\begin{figure}[tbh]
  \centering
  \includegraphics[width=7cm]{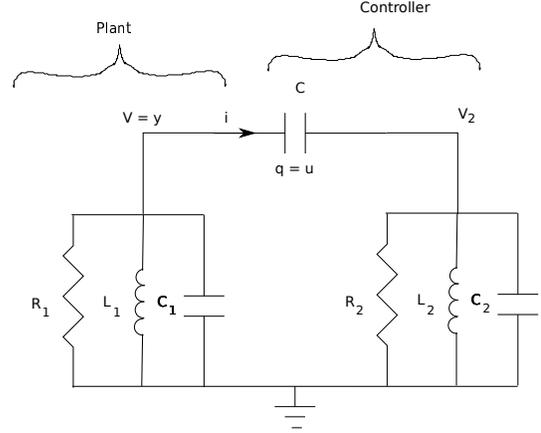}
  \caption{RLC circuit illustration of the NI stability theorem}
  \label{F8}
\end{figure}

We now consider electrical circuit interpretations of the NI stability result Theorem \ref{T1}. We first consider an RLC circuit interpretation of Theorem \ref{T1} as shown in Figure \ref{F8} where the plant input $u$ corresponds to the charge $q = \int i dt$ on the controller coupling capacitor $C$ and the plant output $y$ corresponds to the  voltage $V$  on the plant circuit. 
To find the plant transfer function, we first write down the impedance of the plant circuit:
$
Z_1(s) = \frac{1}{\frac{1}{R_1}+\frac{1}{sL_1}+sC_1}.
$
Then
\[
V = -i Z_1 = -sq Z_1 = - \frac{sq}{\frac{1}{R_1}+\frac{1}{sL_1}+sC_1}.
\]
Hence, the plant transfer function is 
$
\bar G(s) = - \frac{s^2}{s^2C_1+\frac{s}{R_1}+\frac{1}{L_1}}
$
which is SNI. 

Also, to find the controller transfer function, we write down the impedance of the controller circuit:
$
Z_2(s) = \frac{1}{\frac{1}{R_2}+\frac{1}{sL_2}+sC_2}.
$
Then,
\[
V_2 = i Z_2 = sq Z_2 = \frac{sq}{\frac{1}{R_2}+\frac{1}{sL_2}+sC_2}.
\]
Hence,
$
q = C(V-V_2) = CV - \frac{Cs^2q}{s^2C_2 +\frac{s}{R_2}+\frac{1}{L_2}}
$
and therefore
\[
q = \frac{CV}{1+\frac{Cs^2}{s^2C_2 +\frac{s}{R_2}+\frac{1}{L_2}}}
= C\frac{s^2C_2 +\frac{s}{R_2}+\frac{1}{L_2}}{s^2(C+C_2) +\frac{s}{R_2}+\frac{1}{L_2}}V.
\]
That is, 
$
G(s) = C\frac{s^2C_2 +\frac{s}{R_2}+\frac{1}{L_2}}{s^2(C+C_2) +\frac{s}{R_2}+\frac{1}{L_2}}
$
which is NI provided $C \geq 0$, $C+C_2 > 0$, $R_2 > 0$ and $L_2 > 0$. 

 In this case, the DC gain condition (\ref{DC_gain}) of Theorem \ref{T1} is 
$
\bar G(0) G(0) = 0\times C < 1
$
which is automatically satisfied. Thus, in this case, Theorem \ref{T1} can be interpreted as saying that the coupling of the two RLC circuits via a capacitor will lead to the overall system remaining internally stable. In addition, $G(s)$ will retain the NI property even if $C_2$ is negative provided that $C+C_2 > 0$. Thus, in this case, we can also allow for controllers with active elements. 

\begin{figure}[tbh]
  \centering
  \includegraphics[width=7cm]{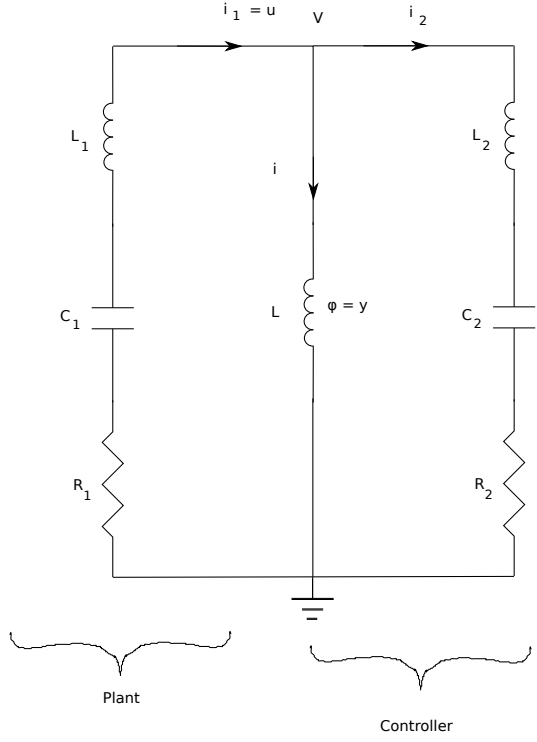}
  \caption{RLC circuit illustration of the NI stability theorem}
  \label{F9}
\end{figure}

An alternative RLC circuit interpretation of Theorem \ref{T1} involves inductive coupling rather than capacitive coupling.  To illustrate this, consider the RLC circuit shown in Figure \ref{F9}. 
In this circuit,  the plant input $u$ corresponds to the flux $\phi = \int V dt$ on the controller coupling inductor and the plant output $y$ corresponds to the  current $i_1$  of the plant circuit. 
To find the plant transfer function, we first write down the impedance of the plant circuit:
$
Z_1(s) = R_1+sL_1+\frac{1}{sC_1}.
$
Then $V = -i_1 Z_1$ implies
\[
i_1 = -\frac{V}{Z_1} = - \frac{s\phi}{R_1+sL_1+\frac{1}{sC_1}}.
\]
Hence, the plant transfer function is 
$
\bar G(s) = - \frac{s^2}{s^2L_1+sR_1+\frac{1}{C_1}}
$
which is SNI provided $L_1 > 0$, $R_1 > 0$ and  $C_1 > 0$. 

Also, to find the controller transfer function, we write down the impedance of the controller circuit:
$
Z_2(s) = R_2+sL_2+\frac{1}{sC_2}.
$
Then,
\[
i_2 =\frac{V}{Z_2} = \frac{s\phi}{R_2+sL_2+\frac{1}{sC_2}}.
\]
Also, $i = \frac{V}{sL} = \frac{\phi}{L}$. Hence,
\begin{eqnarray*}
i_1 &=& i+i_2 = \frac{\phi}{L}+ \frac{s\phi}{R_2+sL_2+\frac{1}{sC_2}}\\
&=& \frac{s^2(L+L_2)+sR_2+\frac{1}{C_2}}{L(s^2L_2+sR_2+\frac{1}{C_2})}\phi
\end{eqnarray*}
and therefore
$
\phi = \frac{L(s^2L_2+sR_2+\frac{1}{C_2})}{L(s^2(L+L_2)+sR_2+\frac{1}{C_2})}i_1
$
That is, 
$
G(s) = \frac{L(s^2L_2+sR_2+\frac{1}{C_2})}{L(s^2(L+L_2)+sR_2+\frac{1}{C_2})}
$
which is NI provided $L \geq 0$, $L+L_2 > 0$, $R_2 > 0$ and $C_2 > 0$. 

 In this case, the DC gain condition (\ref{DC_gain}) of Theorem \ref{T1} is 
\[
\bar G(0) G(0) = 0\times L < 1
\]
which is automatically satisfied. Thus, in this case, Theorem \ref{T1} can be interpreted as saying that the coupling of the two RLC circuits via an inductor will lead to the overall system remaining internally stable. 

\section{Conclusions}
\label{sec:Conclusions}
In this paper we have presented a number of physical interpretations of stability results for  feedback interconnections of negative imaginary systems and compared these with stability results for positive real systems. The physical interpretations have involved both spring mass damper systems and RLC electrical networks. The results may be useful in gaining a better understanding of these stability results and motivating new extensions to the existing results. Also, the interpretations of a controller as a physical system may be useful in robust controller design and tuning even in the case when the controller is actually implemented using actuators, sensors and computers; e.g, see \cite{MOH03}. 

\section*{Acknowledgement}
The starting point for the physical interpretations presented in this paper was a discussion with Martin Corless. 

% \bibliography{/home/irp/Bibliog/irpnew}

\begin{thebibliography}{10}
\providecommand{\url}[1]{#1}
\csname url@rmstyle\endcsname
\providecommand{\newblock}{\relax}
\providecommand{\bibinfo}[2]{#2}
\providecommand\BIBentrySTDinterwordspacing{\spaceskip=0pt\relax}
\providecommand\BIBentryALTinterwordstretchfactor{4}
\providecommand\BIBentryALTinterwordspacing{\spaceskip=\fontdimen2\font plus
\BIBentryALTinterwordstretchfactor\fontdimen3\font minus
  \fontdimen4\font\relax}
\providecommand\BIBforeignlanguage[2]{{%
\expandafter\ifx\csname l@#1\endcsname\relax
\typeout{** WARNING: IEEEtran.bst: No hyphenation pattern has been}%
\typeout{** loaded for the language `#1'. Using the pattern for}%
\typeout{** the default language instead.}%
\else
\language=\csname l@#1\endcsname
\fi
#2}}

\bibitem{LP06}
A.~Lanzon and I.~R. Petersen, ``Stability robustness of a feedback
  interconnection of systems with negative imaginary frequency response,''
  \emph{IEEE Transactions on Automatic Control}, vol.~53, no.~4, pp.
  1042--1046, 2008, arXiv:1401.7739.

\bibitem{LP_CSM}
I.~R. Petersen and A.~Lanzon, ``Feedback control of negative-imaginary
  systems,'' \emph{Control Systems Magazine}, vol.~30, no.~5, pp. 54 -- 72,
  2010, arXiv:1401.7745.

\bibitem{ANG06}
D.~Angeli, ``Systems with counterclockwise input-output dynamics,'' \emph{IEEE
  Transactions on Automatic Control}, vol.~51, no.~7, pp. 1130--1143, 2006.

\bibitem{BLME07}
B.~Brogliato, R.~Lozano, B.~Maschke, and O.~Egeland, \emph{Dissipative Systems
  Analysis and Control: Theory and Applications}, 2nd~ed.\hskip 1em plus 0.5em
  minus 0.4em\relax New York: Springer-Verlag, 2007.

\bibitem{XiPL1}
J.~Xiong, I.~R. Petersen, and A.~Lanzon, ``A negative imaginary lemma and the
  stability of interconnections of linear negative imaginary systems,''
  \emph{IEEE Transactions on Automatic Control}, vol.~55, no.~10, pp. 2342 --
  2347, 2010.

\bibitem{MKPL10}
M.~A. Mabrok, A.~G. Kallapur, I.~R. Petersen, and A.~Lanzon, ``Generalizing
  negative imaginary systems theory to include free body dynamics: Control of
  highly resonant structures with free body motion,'' \emph{IEEE Transactions
  on Automatic Control}, vol.~59, no.~10, pp. 2692--2707, October 2014,
  arXiv:1305.1079.

\bibitem{PW98}
J.~W. Polderman and J.~C. Willems, \emph{Introduction to Mathematical Systems
  Theory: A Behavioral Approach}.\hskip 1em plus 0.5em minus 0.4em\relax New
  York: Springer, 1998.

\bibitem{AV73}
B.~D.~O. Anderson and S.~Vongpanitlerd, \emph{Network Analysis and
  Synthesis}.\hskip 1em plus 0.5em minus 0.4em\relax Mineola, NY: Dover
  Publications, 2006.

\bibitem{PRE11}
A.~Preumont, \emph{Vibration Control of Active Structures}, 3rd~ed.\hskip 1em
  plus 0.5em minus 0.4em\relax Berlin: Springer-Verlag, 2011.

\bibitem{BAL79}
M.~Balas, ``Direct velocity feedback control of large space structures,''
  \emph{Journal of Guidance and Control}, vol.~2, no.~3, pp. 252--253, 1979.

\bibitem{MOH03}
S.~Moheimani, ``A survey of recent innovations in vibration damping and control
  using shunted piezoelectric transducers,'' \emph{IEEE Transactions on Control
  System Technology}, vol.~11, no.~4, pp. 482--494, 2003.

\end{thebibliography}
% \bibliographystyle{IEEEtran}

\end{document}